\documentclass[aps,prl,twocolumn,showpacs,preprintnumbers,amsmath,amssymb]{revtex4-1}%
\usepackage{graphicx}
\usepackage{dcolumn}
\usepackage{epsfig,bm}
\usepackage{amsfonts}
\usepackage{color}
\usepackage{bigints}

\begin{document}

\title{The role of Mie scattering in the seeding of matter-wave superradiance}

\author{R. Bachelard$^{a}$}
\author{H. Bender$^{a,c}$}
\author{Ph.W. Courteille$^{a}$}
\affiliation{$^{a}$Instituto de F\'isica de S\~ao Carlos, Universidade de S\~ao Paulo, 13560-970 S\~ao Carlos, SP, Brazil}
\author{N. Piovella$^{b}$}
\affiliation{$^{b}$Dipartimento di Fisica, Universit\`a Degli Studi di Milano and INFM,Via Celoria 16, I-20133 Milano, Italy}
\author{C. Stehle$^{c}$}
\author{C. Zimmermann$^{c}$}
\author{S. Slama$^{c}$}
\affiliation{$^{c}$Physikalisches Institut, Eberhardt-Karls-Universit\"at T\"ubingen, D-72076 T\"ubingen, Germany}

\date{\today}

\begin{abstract}
Matter-wave superradiance is based on the interplay between ultracold
atoms coherently organized in momentum space and a backscattered wave.
Here, we show that this mechanism may be triggered by Mie scattering
from the atomic cloud. We show how the laser light populates the modes of the cloud, and thus imprints a phase gradient on the excited atomic dipoles. The interference with the atoms in the ground state results in a grating that in turn generates coherent emission, contributing to the backward light wave onset. The atomic recoil 'halos' created by the Mie-scattered light exhibit a strong anisotropy, in contrast to single-atom scattering.
\end{abstract}


\maketitle

Matter wave superradiance (MWSR) \cite{Inouye99} and collective atomic recoil lasing (CARL) \cite{Bonifacio94,Kruse03} are light-induced instabilities of the density distribution in atomic clouds. More precisely, they are due to correlations between successive scattering events mediated by long-lived coherences in the motional state of an (ultracold) atomic cloud or in the light field of an optical resonator \cite{Slama07}. Despite considerable theoretical efforts having been devoted to the dynamics of MWSR \cite{Piovella97,Piovella01,Ketterle01}, open questions still remain. One of them concerns the seeding mechanism which is able to start the MWSR instability even in the presence of losses. Thermal and quantum fluctuations will naturally contribute. However, in this paper we point out the particular role of Mie scattering, which turns out to be important at the onset of MWSR. Indeed, Mie scattering is active before any instabilities have developed, and it induces a {\it phase} correlation between the atomic dipoles that favors the build up of an instability. 

The below-threshold dynamics and the seeding of matter-wave superradiance are interesting problems. As long as we consider the atomic cloud as a homogeneous entity, e.g.,~a Bose-Einstein condensate (BEC) in the mean field description, no scattering should occur at all. Theoretical models which describe BEC's as matter waves without fluctuations thus fail to explain how MWSR is initiated in the absence of a seeding wave~\cite{Kozuma99}. In contrast, recent work has shown \cite{Bienaime10,Bender10} how atomic coarse-graining, density fluctuations and Mie scattering from finite-sized clouds can even influence the scattering of a single photon. Also optical cavities may strongly  affect the scattering by shaping the angular distribution of the density of modes that are capable of receiving the scattered photons~\cite{Bux11}. These processes have a decisive impact on the mode competition preceding the exponential instability, and thereby on the instability itself.

Here we show that Mie scattering, caused by the finite size of the atomic cloud, favors the formation of a matter wave-dipole grating. This has already been suspected in \cite{Ketterle10}. Indeed, prior to any significant motion of the atoms, their dipoles collectively order, which in turn leads to coherent emission.

In previous papers \cite{Courteille10,Bux10,Bachelard11,Bachelard12}, we discussed the impact of atomic coarse-graining and finite scattering volumes on the radiation pressure force which acts on the cloud's center of mass. Here, we investigate the momentum distribution following the cooperative scattering of the laser light by a BEC. Our theoretical model describes the atomic cloud as a macroscopic matter wave that is homogeneously distributed within a sphere, i.e., the atoms are considered to be strongly delocalized and density fluctuations are neglected. We find that the momentum distribution of the atoms adopts the shape of a {\bf recoil halo}, very similar to the ones observed experimentally in time-of-flight images of BEC's. The halo indicates the directions into which the atoms are preferentially scattered before the density distribution is noticeably modified. In particular, it exhibits a pronounced peak at $2\hbar k$. This corresponds to an increased backscattering of light that acts as a seed for MWSR.

We consider an homogeneous spherical cloud of two-level atoms illuminated by a laser of wavevector $\mathbf{k_0}=k_0\hat{z}$, Rabi frequency $\Omega_0=d E_0/\hbar$ (where $d$ is the electric dipole matrix element and $E_0$ the laser electric field) and detuned from the atomic transition by $\Delta_0$.
The atomic cloud is described as a bosonic ensemble of $N$ two-level atoms with field operator $\hat\Psi(\mathbf{r},t)=\hat\Psi_g(\mathbf{r},t)+\hat\Psi_e(\mathbf{r},t)$
($g$ for the ground state, $e$ for the excited one). We treat the condensate as an ideal gas and consider the scattering between matter waves and optical waves, but we neglect nonlinearities due to atom-atom interaction. 
In second quantization, the interaction between the atoms and light is described by the Hamiltonian~\cite{Courteille10,Bienaime11}:
\begin{eqnarray}\label{HQ}
    \hat H(t)&=&\frac{\hbar\Omega_0}{2} \int d\mathbf{r} \left[\hat\Psi^\dagger_{e}(\mathbf{r},t)\hat\Psi_{g}
    (\mathbf{r},t)e^{-i\Delta_0 t+i\mathbf{k}_0\cdot \mathbf{r}}+h.c. \right]    
    \\ &+&\hbar\sum_{\mathbf{k}}g_{\mathbf{k}}
    \int d\mathbf{r} \left[\hat\Psi^\dagger_{e}(\mathbf{r},t)\hat\Psi_{g}
    (\mathbf{r},t)\hat a_{\mathbf{k}} e^{-i\Delta_k t+i\mathbf{k}\cdot
    \mathbf{r}}+h.c.\nonumber
\right],
\end{eqnarray}
where $g_k=d[\omega_k/(2\hbar\epsilon_0 V_v)]^{1/2}$ and
$V_v$ is the quantization volume. In Eq.(\ref{HQ}), the first (second) line describes
the absorption and emission of a pump mode $\Omega_0$ (a vacuum mode
$\hat a_{\mathbf{k}}$). Doppler effects are neglected.
Replacing $\hat\Psi_{e}(\mathbf{r},t)\rightarrow\hat\Psi_{e}(\mathbf{r},t)e^{i\Delta_0 t}$ induces an energy shift of $-\Delta_0|\hat\Psi_{e}|^2$ in the Hamiltonian, from which the Heisenberg equations may be derived:
\begin{eqnarray}
  \frac{\partial\hat\Psi_g}{\partial t} &=& -i\hat\Psi_{e}\Bigg[\frac{\Omega_0}{2}e^{-i\mathbf{k}_0\cdot\mathbf{r}}+\sum_{\mathbf{k}}g_{\mathbf{k}} \hat a_{\mathbf{k}}^\dagger e^{-i(\Delta_0-\Delta_k)t-i\mathbf{k}\cdot \mathbf{r}}\Bigg],\label{eq:psig}
   \\    \frac{\partial\hat\Psi_e}{\partial t} &=& -i\hat\Psi_{g}\Bigg[\frac{\Omega_0}{2}e^{i\mathbf{k}_0\cdot\mathbf{r}}+\sum_{\mathbf{k}}g_{\mathbf{k}}  \hat a_{\mathbf{k}} e^{i(\Delta_0-\Delta_k)t+i\mathbf{k}\cdot \mathbf{r}}\Bigg]\nonumber
   \\ && +i\Delta_0\hat\Psi_e \label{eq:psie}
 \\  \frac{\mbox{d} \hat a_\mathbf{k}}{\mbox{d} t} &=& -ig_{\mathbf{k}}
    e^{-i(\Delta_0-\Delta_k)t}
    \int d\mathbf{r}
    \hat\Psi^\dagger_{g}(\mathbf{r},t)\hat\Psi_{e}
    (\mathbf{r},t) e^{-i\mathbf{k}\cdot
    \mathbf{r}}.\label{eq:ak}
\end{eqnarray}
For large atom numbers and far detuning from the atomic transition frequency, one can neglect quantum fluctuations and treat the operators as c-numbers ($\hat\Psi\rightarrow\psi$, $\hat a_\mathbf{k}\rightarrow a_\mathbf{k}$).
Eq.(\ref{eq:ak}) is integrated over time as
\begin{equation}
a_\mathbf{k}=-ig_\mathbf{k} \int_0^t \mbox{d} t'e^{-i(\Delta_0-\Delta_k)t'}
    \int d\mathbf{r}\psi^*_{g}(\mathbf{r},t')\psi_{e}(\mathbf{r},t) e^{-i\mathbf{k}\cdot \mathbf{r}},\label{eq:ak2}
\end{equation}
and inserted into Eq.(\ref{eq:psie}). Furthermore, we  
switch to a continuous-mode description $\sum_\mathbf{k}\to (V_\nu/(2\pi)^3)\int\mbox{d}\mathbf{k}$ and obtain: 
\begin{eqnarray}
&&\frac{\partial\psi_e}{\partial t}  =   i\Delta_0\psi_e(\mathbf{r},t)-i\frac{\Omega_0}{2}e^{i\mathbf{k}_0.\mathbf{r}}\psi_{g}(\mathbf{r},t)-\psi_{g}(\mathbf{r},t)\int \mbox{d}\mathbf{r}' \label{eq:psie2}
   \\ &&\times \int\mbox{d}\mathbf{k}g_{\mathbf{k}}^2
e^{i\mathbf{k}\cdot(\mathbf{r}-\mathbf{r}')} \int_0^t\mbox{d} t' e^{i(\Delta_0-\Delta_k)(t-t')} \psi_{g}^*(\mathbf{r}',t)\psi_{e}(\mathbf{r}',t').\nonumber
\end{eqnarray}
We consider time scales over which the atomic density does not significantly change, i.e., $\psi_g(\mathbf{r},t)\approx\psi_{g0}(\mathbf{r})$ in Eq.(\ref{eq:psie2}), with $\rho_0(\mathbf{r})=|\psi_{g0}(\mathbf{r})|^2$ the initial density of the cloud. Using the Markov approximation, i.e., the photon time of flight through the cloud is much shorter than the atomic decay time, the last integral in Eq.(\ref{eq:psie2}) is replaced by $\delta(k-k_0)\psi_{g0}(\mathbf{r}')\psi_e(\mathbf{r}',t)/c$. With the assumption that all the electromagnetic modes are equally present in the system ($g_\mathbf{k}\approx g_{k_0}$) and by keeping rotating-wave-approximation terms, one can show that~\cite{Svidzinsky10}
\begin{equation}
\int\mbox{d}\mathbf{k} g_{\mathbf{k}}^2 e^{i\mathbf{k}\cdot\mathbf{d}}\int_0^\infty\mbox{d} t' e^{i(\Delta_0-\Delta_k)(t-t')}=\frac{\Gamma}{2ik_0|\mathbf{d}|}e^{ik_0|\mathbf{d}|},
\end{equation}
where $\Gamma=V_\nu g_{k_0}^2k_0^2/\pi c=d^2k_0^3/(2\pi\hbar\epsilon_0)$ is the atomic decay rate.
For the normalized excitation field $\beta(\mathbf{r},t)=\psi_e(\mathbf{r},t)/\psi_{g0}(\mathbf{r})$, with $|\beta(\mathbf{r})|^2$ the probability for an atom to be excited, one obtains~\footnote{The single-atom spontaneous decay term $-(\Gamma/2)\beta$ in
Eq.(\ref{eq:beta})  arises from the non-commutative properties of the bosonic operators $\hat\Psi_g$ and $\hat\Psi_e$, analogously to the spontaneous emission from vacuum arising
from the commutation rule for $\hat a_\mathbf{k}$. The single-atom decay term, however, is negligible far from
the atomic transition ($\Delta_0\gg\Gamma$), as will be assumed in the present work.}:
\begin{eqnarray}
\frac{\partial\beta(\mathbf{r},t)}{\partial t} &=&
   \left(i\Delta_0-\frac{\Gamma}{2}\right)\beta(\mathbf{r},t)-i\frac{\Omega_0}{2}e^{i\mathbf{k}_0\cdot\mathbf{r}}\nonumber
   \\ &-&\frac{\Gamma}{2}\int \mbox{d}\mathbf{r}'\rho_0(\mathbf{r}')\frac{\exp\left(ik_0|\mathbf{r}-\mathbf{r}'|\right)}{ik_0|\mathbf{r}-\mathbf{r}'|}\beta(\mathbf{r}',t),
\label{eq:beta}
\end{eqnarray}
which recovers the model of cooperative scattering of a plane-wave introduced in~\cite{Courteille10}.

In the steady-state regime, Eq.(\ref{eq:beta}) can be used to describe the scattering of light from a dielectric medium with an index of refraction~\cite{Bachelard12}
\begin{equation}
m_c=\sqrt{1-4\pi\rho_0\Gamma/k_0^3(2\Delta_0+i\Gamma)}.
\end{equation}
In this regime of linear optics, the field $\beta$ is directly proportional to the electric field, and the ratio between both is the susceptibility. Hence, the excitation pattern inside the cloud can be calculated analogously to Mie's theory~\cite{Mie1908,Bohren98}. The polarization amplitude $\beta$ inside the cloud is decomposed into
modes, thereafter labeled $n$, which are elementary solution of the Helmholtz equation:
\begin{equation}
(\Delta+m_c^2k_0^2)\beta=0.\label{eq:Helmholtz}
\end{equation}
In spherical homogeneous distributions (with coordinates $\mathbf{r}(r,\theta,\phi)$,
where $\theta$ is the angle with respect to the $\hat z$ axis) these solutions are of the form $j_n(m_ck_0r)P_n(\cos\theta)$ inside the cloud, with $j_n$ the spherical Bessel function and $P_n$ the Legendre polynomials. Then
the amplitude $\beta_n$ of each mode are calculated according to the
method described in Ref.~\cite{Bachelard12}, giving for an uniform spherical cloud:
\begin{equation}
\beta(\mathbf{r})=\frac{\Omega_0}{\Gamma}\sum_{n=0}^\infty (2n+1)i^n\beta_n j_n(m_ck_0r)P_n(\cos\theta),\label{eq:betan}
\end{equation}
where the coefficients $\beta_n$ are given by
\begin{equation}
\beta_n=\frac{j_n(k_0R)}{(2\delta+i)j_n(m_ck_0R)+i\lambda_n h_n^{(1)}(k_0R)},
\end{equation}
with $R$ the radius of the sphere and $h_n^{(1)}$ the spherical Hankel function. In general, the radiation is accurately described by considering a number of modes on the order of the size of the system $k_0R$~\cite{Bohren98}. Note though that such a coarse-grained approach to describe the cloud cannot capture features associated with disorder (e.g., Anderson localization).

As can be deduced from Eq.(\ref{eq:Helmholtz}), the laser imprints a phase with wavevector $m_ck_0$ to the excited state $\psi_e$, which interferes with the ground state $\psi_g$ that has a constant phase.
The resultant grating is at the origin of coherent emission and of the backscattering wave that acts as a seed for MWSR. Indeed, the field radiated by the atoms inside the cloud in a direction $\mathbf{u}(\theta,\phi)$ is proportional to the structure factor:
\begin{equation}
s_c(\mathbf{u})=\frac{1}{N}\int\rho(\mathbf{r})\beta(\mathbf{r})e^{-im_ck_0\mathbf{u}\cdot\mathbf{r}}\mbox{d}\mathbf{r},\label{eq:sc}
\end{equation}
with the index of refraction $m_c$ given above. Hence, the coherent emission by the cloud is a direct consequence of the periodic excitation field and the resulting grating. Note that we do not perform an adiabatic elimination of the excited state~\cite{Gardiner00}; we rather extract the scattering pattern from this state.

The scattered light is difficult to observe directly. However, the radiation pattern is also present in the momentum distribution of the atoms, which can be easily recorded by time-of-flight imaging. Different from an $N$-body model, the quantum matter field approach yields the momentum distribution simply through the Fourier transform of the matter field, $\widehat{\psi}(\mathbf{p})=(2\pi\hbar)^{-3}\int \psi(\mathbf{r})e^{-i\mathbf{p}\cdot\mathbf{r}/\hbar}\mbox{d}\mathbf{r}$. Hence, for an homogeneous cloud, the momentum distribution of the excited state is directly proportional to the structure factor: $\widehat\psi_e(\mathbf{p})\propto s_c(\mathbf{p}/\hbar)$, with $\mathbf{p}=m_c\hbar\mathbf{k}$ and $\mathbf{k}=k_0\mathbf{u}$. Using Eq.(\ref{eq:betan}), it can be deduced that the momentum wavefunction $\widehat\psi_e$ for an uniform sphere of radius $R$ reads:
\begin{equation}
\widehat\psi_e(\mathbf{p})=\frac{\Omega_0\sqrt{\rho_0}}{\Gamma(2\pi\hbar)^3}\sum_{n=0}^\infty(2n+1)\beta_n\gamma_n(p)P_n(\cos\theta),\label{eq:hatpsie}
\end{equation}
with $\gamma_n(p)=4\pi\int_0^R r^2j_n(m_ck_0r)j_n(pr/\hbar)\mbox{d} r$.
\begin{figure}[!ht]
\centering
\begin{tabular}{c}
\epsfig{file=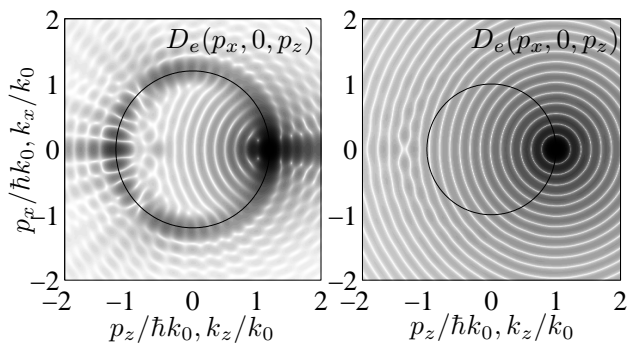,width=8.3cm}
\end{tabular}
\caption{Momentum distribution of the excited state $D_e=|\widehat\psi_e|^2$ in logarithmic scale. The emission pattern $|s_c(\mathbf{k})|^2$ concentrates around a circle with radius $p=m_c\hbar k_0$ (black line). Simulations for a laser detuning $\Delta_0=-3$GHz and a cloud (a) of size $k_0R=20$, atom number $N=1.15\times10^6$ and refractive index $m_c=1.2$, (b) of size $k_0R=20$, atom number $N=100$ and refractive index $m_c=1+2.10^{-7}$ ($\Delta_0=-488$MHz). For rubidium, $k_0^{Rb}=8.05\times10^6$m$^{-1}$ and $\Gamma^{Rb}=6.1$MHz.\label{fig:psie}}
\end{figure}
For a very small cloud ($k_0R\ll 1$), the scattering of light is isotropic, as expected from Rayleigh theory: only the first Mie mode ($n=0$) is populated. However, for large many-particle spheres, Mie scattering turns out to be fundamentally anisotropic, as many modes are populated.
Fig.\ref{fig:psie}(a) shows a typical momentum distribution of the excited atoms $|\widehat\psi_e(\mathbf{p})|^2$ and the associated scattering pattern $|s_c(\mathbf{k})|^2$ of the light. Although most atoms recoil to $\mathbf{k}=k_0\hat z$, a significant amount of light is scattered backward ($\mathbf{k}=-k_0\hat z$) and acts as a seed for the MWSR instability.

Note that the momentum distribution $D_e(\mathbf{p})=|\widehat\psi_e(\mathbf{p})|^2$ is concentrated along a circle with radius $p=m_c\hbar k_0$ rather than $\hbar k_0$ (i.e., that $\gamma_n(p)$ reaches a maximum for $p=m_c\hbar k_0$). This is a signature of the Minkowski momentum for atomic recoil, that characterizes the momentum exchanged between light and matter in dielectric media~\cite{Campbell05}. The blurring of the momentum wavefunction along the circle originates in the finite size of the cloud that creates a natural momentum spread $\sigma_p\sim\hbar/R$. The ripples of the distribution are due to the sharp boundary of the cloud's density, yielding a Fourier transform with many secondary peaks.

In optical dilute clouds, almost all of the light is scattered forward, leaving the incoming light almost untouched. Such a scattering pattern is displayed in Fig.\ref{fig:psie}(b), where the intricate pattern of Mie scattering by an optically dense cloud has disappeared. In this limit, the excitation field is often approximated by the timed Dicke state $\beta(\mathbf{r})\sim\beta_{TDS}e^{i\mathbf{k}_0\cdot\mathbf{r}}$~\cite{Scully06}, yet such an ansatz fails to predict any three-dimensional recoil pattern. 

\begin{figure}[!ht]
\centering
\begin{tabular}{cc}
\epsfig{file=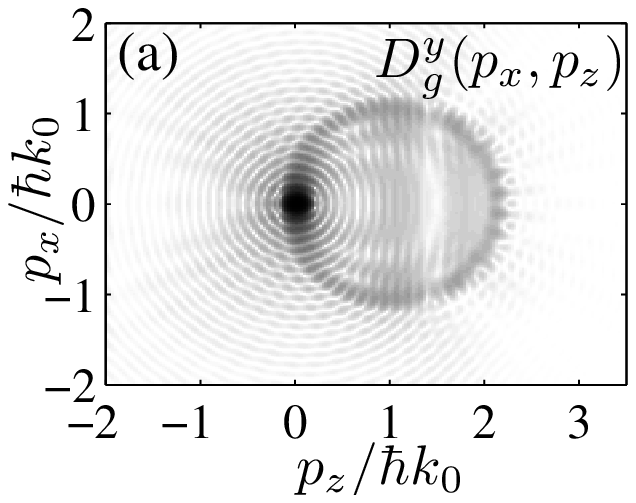,height=2.9cm}&\epsfig{file=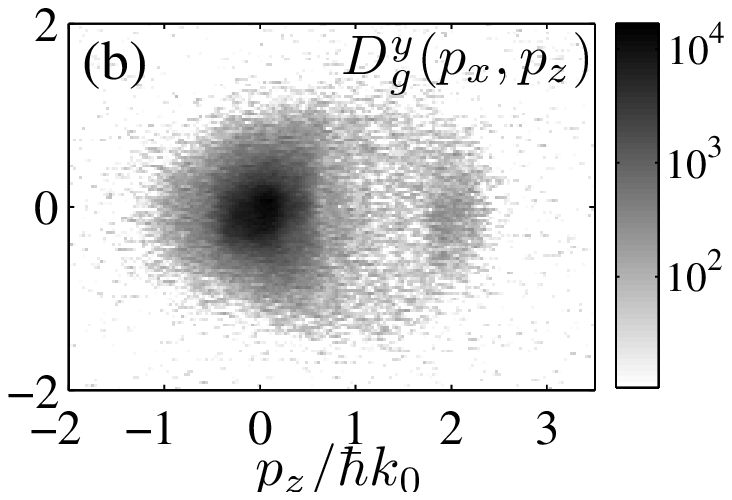,height=2.9cm}
\end{tabular}
\caption{(a) Integrated momentum distribution of the ground state, calculated from Eq.\ref{eq:hatpsie} for a spherical homogeneous cloud of radius $k_0R=29.6$, $\Delta_0=-15$GHz and with a refractive index $m_c=1.067$ (a low-pass Gaussian filter was applied to attenuate the ripples due to the sharp boundaries of the homogeneous clouds, since they are irrelevant for comparison with the experiment) (b) Experimental integrated momentum distribution of the ground state for an ellipsoidal cloud of length $k_0\sigma_z\sim29.6$, transverse radius $k_0\sigma_\perp\sim4.7$, with $N\sim 147000$ atoms and a laser detuning $\Delta_0=-15$GHz and a $20\mu s$ laser pulse of $17$mW. Its refractive index is $m_c\approx1.067$.\label{fig:expthr}}
\end{figure}

After an atom absorbs a photon from the laser with a momentum kick $m_c\hbar\mathbf{k}_0$, the photon is reemited according to the Mie pattern $s_c(\mathbf{p}/\hbar)\propto \widehat\psi_e(\mathbf{p})$. Thus the atom will gain an extra momentum $m_c\hbar k_0$ with a direction opposite to the emitted photon, and the momentum pattern of the ground state atoms after the scattering process is given by  $|\widehat{\psi}_e(m_c\hbar \mathbf{k}_0-\mathbf{p})|^2$. Experimentally, the column-integrated momentum distribution is observed in time-of-flight images. This leads to defining the projected distribution $D_g^y(p_x,p_z)=\int|\hat\psi_g(\mathbf{p})|^2\mbox{d} p_y$. Such an integrated distribution is presented in Fig.\ref{fig:expthr}(a). The atoms are observed to {\it inhomogeneously} fill a circle of radius $m_c\hbar k_0$. In particular, the part of the sphere around $\mathbf{p}=(3/2)\hbar\mathbf{k}_0$ is weakly populated, which reflects the anisotropic nature of Mie scattering.

Experimentally, the atomic recoil patterns are investigated by using the set-up of~\cite{Bender10}. After the interaction with the light, the $^{87}$Rb atoms ballistically expand for $t_F=20ms$. Since the initial size of the cloud $\sim 20\mu m$ is much smaller than its expanded size at the time of imaging, the density distribution of the expanded cloud is recorded by standard absorption imaging yielding the initial momentum distribution before the expansion according to $\mathbf{p}=m_{Rb}\mathbf{r}/t_F$.

The integrated momentum distribution observed experimentally reproduces the features predicted by Mie scattering (see Fig.\ref{fig:expthr}(b)). The sphere of radius $\hbar k_0$ is filled with atoms, yet it exhibits a region where the probability of the atomic recoil is very low around $p_z=1.5\hbar k_0$. Moreover, it can be observed that a large number of atoms recoil around $\mathbf{p}=2\hbar\mathbf{k}_0$, which is the signature of the backward emitted wave. The presence of preferred directions of emission is associated with the presence of Mie resonances~\cite{Bachelard12}: the increase of emission in these directions can be interpreted as a self-Purcell enhancement~\cite{Purcell46} because the cloud acts as a cavity on itself to modify its emission. This is in contrast to single-atom scattering, where light is scattered in random directions and not specifically backward and forward.

It is important to note that Mie scattering is a three-dimensional process, while MWSR only develops along its most unstable direction. Mie scattering is a seeding process that emits light in many directions; including backwards, which is known to be the most unstable direction for MWSR in a cigar-shaped cloud illuminated along its main axis~\cite{Moore99}.

\begin{figure}[!ht]
\centering
\begin{tabular}{c}
\epsfig{file=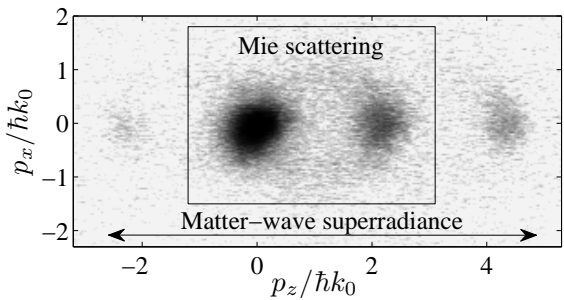,width=8cm}
\end{tabular}
\caption{Momentum distribution $|\widehat\psi_g|^2$ {\it above} the threshold of the MWSR. Experiment realized with an ellipsoidal cloud of transverse radius $k_0\sigma_\perp\sim3.5$, length $k_0\sigma_z\sim22$, with $N\sim 156000$ atoms and a laser detuning $\Delta_0=-15$GHz.\label{fig:CARL}}
\end{figure}

These two processes are illustrated in Fig.\ref{fig:CARL}. Initially, Mie scattering populates a sphere of radius $p\approx m_c\hbar k_0$. Then a MWSR instability develops, i.e., the grating induces light emission that, in turn, amplifies this atomic grating. The atoms observed in $\mathbf{p}\approx -2m_c\hbar \mathbf{k}_0$ and $\mathbf{p}\approx4m_c\hbar \mathbf{k}_0$ are not predicted by Mie scattering and can be explained only by the self-consistent matter-wave dynamics.

In this paper we showed that Mie scattering induces a grating in the atomic distribution even below the threshold for the MWSR instability. Its  signature is an anisotropic three-dimensional halo in the atomic momentum distribution. The atoms observed at $\mathbf{p}\approx 2\hbar \mathbf{k}_0$ generate a seeding wave for MWSR. Indeed, the matter wave modes at $0$ and $2\hbar \mathbf{k}_0$ together form a density grating, at which subsequent light injected from the pump laser is Bragg-scatterered in a self-amplifying process.

Note that quantum fluctuations and disorder may also give rise to backward emission~\cite{Wilkowski04}, and thus act as a seed for the matter-wave superradiance. These effects will be included in an extended work.

Finally, it is interesting to remark that the off-axis emission of photons should be associated with higher modes ($n\gg1$) that correspond to photons with long lifetime within the cloud~\cite{Bienaime12}. Thus, a time-resolved observation of the off-axis atomic recoils should bear the signature of subradiance.

We acknowledge helpful discussion with R. Kaiser. This work has been supported by the Funda\c{c}\~ao de Amparo à Pesquisa do Estado de S\~ao Paulo (FAPESP) and the Research Executive Agency (program COSCALI, No. PIRSES-GA-2010-
268717). C.S., C.Z and S.S acknowledge support by the Deutsche Forschungsgemeinschaft.

\end{document}